\documentclass{appolb}
\usepackage{epsfig}

\begin{document}
\title{THE LIFETIME
OF THE $\Lambda$ - HYPERON BOUND IN HYPERNUCLEI PRODUCED BY p+U COLLISIONS%
}
\author{P. Kulessa$^{a,b}$,
W. Cassing$^{c}$,
L. Jarczyk$^{d}$,
B. Kamys$^{d}$,
H. Ohm$^{a}$,\\
K. Pysz$^{a,b}$,
Z. Rudy$^{d}$,
H. Str\"oher$^{a}$
\address{$^a$Institut f\"ur Kernphysik, Forschungszentrum J\"ulich,
D-52425 J\"ulich, Germany \\
$^b$H. Niewodnicza\'nski Institute of Nuclear Physics,
PL-31342 Cracow, Poland\\
$^c$Institut f\"ur Theoretische Physik,
Justus Liebig Universit\"at Giessen,\\
D-35392 Giessen, Germany\\
$^d$M. Smoluchowski Institute of Physics, Jagellonian University,
PL-30059 Cracow, Poland     }
}

\maketitle
\begin{abstract}
The nonmesonic decay of the $\Lambda$--hyperon has been investigated
by observation of delayed fission from heavy hypernuclei produced in
proton -- U collisions at T$_p$=1.9 GeV.
The lifetime of heavy hypernuclei with
masses A$\approx$220  obtained in the present work, i.e.
$\tau_{\Lambda}= [138 \pm 6 (stat.) \pm 17 (syst.)$] ps,
is the most accurate result for heavy hypernuclei produced in proton
and antiproton induced collisions on a U target so far.
\end{abstract}
\PACS{13.30.-a, 13.75.Ev, 21.80, 25.80.Pw}

\section{Introduction}

$\Lambda$-hyperons bound in hypernuclei do not decay in the same
way as  free hyperons because the mesonic
decay $\Lambda \rightarrow \pi+N$
is strongly Pauli blocked for all but the lightest hypernuclei.
This is due to the fact that
the emerging nucleons
%
have an energy smaller than the Fermi energy of the
nucleons in the hypernucleus.
However, another type of weak hyperon decays
becomes possible in the nuclear medium due to the presence of nucleons,
i.e. the processes
$\Lambda$+N $\rightarrow$ N+N, or
$\Lambda$+N+N $\rightarrow$ N+N+N.  The energy of the final nucleons from
these reactions
is much higher than the Fermi energy and thus this
"nonmesonic decay" is not Pauli blocked.  The study of the nonmesonic decay,
which proceeds via a weak interaction only (the Coulomb and strong
interactions preserve the strangeness), enables one to obtain information
on the weak baryon - baryon interaction which is not accessible
by other means.
Furthermore,
the nonmesonic decay allows to observe strangeness changing -- $\Delta$S=1
weak decays -- whereas the weak nucleon-nucleon interation is limited to
$\Delta$S=0 processes.

The nonmesonic decay of $\Lambda$ hyperons in heavy hypernuclei is
very interesting because of two reasons: i) it corresponds to the decay of the
hyperon in infinite nuclear matter (the hyperon decays from the
S-state for which
the wave function is well localised in the center of the
hypernucleus and surface
effects are negligible), and ii) the mesonic decay in heavy hypernuclei
is completely negligible.
In spite of this
 the available experimental data for the  decay of $\Lambda$ hyperons
in heavy nuclei are scarce and have large uncertainties, which is
essentially
due to the immense difficulty in producing $\Lambda$ hypernuclei and
subsequently detecting their decay products.
Most of the measurements have been performed for
light hypernuclei, i.e. lighter than $_{\Lambda}Fe$
(see {\it e.g.} the review article
\cite{COH90A} or refs. \cite{SZY91,NOU95,BHA98}).
An investigation  of heavy hypernuclei decays
 has been up to now performed only using
Au \cite{KAM01}, Bi \cite{KUL98}, and U \cite{OHM97,PYS98,ZYC99,KUL00} targets
in proton
or antiproton induced reactions \cite{BOC86,ARM93}
or  -- in case of the Bi target -- by electrons \cite{NOG86} .

In the present work we report on an experiment devoted to the measurement
of the lifetime of heavy hypernuclei as produced in proton collisions with
an uranium target.  This experiment was performed with the aim to achieve
an accuracy
comparable to that obtained for heavy hypernuclei produced with Au and Bi
targets. The outline of the paper is as follows:
In section 2  the general concept of the experimental
setup and the data analysis is presented.
%
The third section is devoted to a discussion
of systematic errors arising from various sources while  section 4 contains a comparison of
our present results with the data from the literature as well as a summary.

\section{Experimental setup and data analysis}
\label{sec:exper}
The lifetime of $\Lambda$ hyperons in heavy hypernuclei may be extracted
from the delayed fission channel of heavy hypernuclei, {\it i.e.} fission induced
by the decay of a  hyperon via the process $\Lambda N \rightarrow NN$.
Due to very different time scales for the fission
($\sim 10^{-18}$ s) and for the decay of hyperons ($\sim 2 \cdot 10^{-10}$ s) it can be assumed
 that the delayed
fission occurs almost immediately after the hyperon decay.
Thus the
position distribution of
delayed fission  events from hypernuclei, that move with known velocities in beam direction,
provides unambiguous information on the lifetime distribution of the hyperon
decays.

\begin{figure}[h]
\centering
\epsfig{file=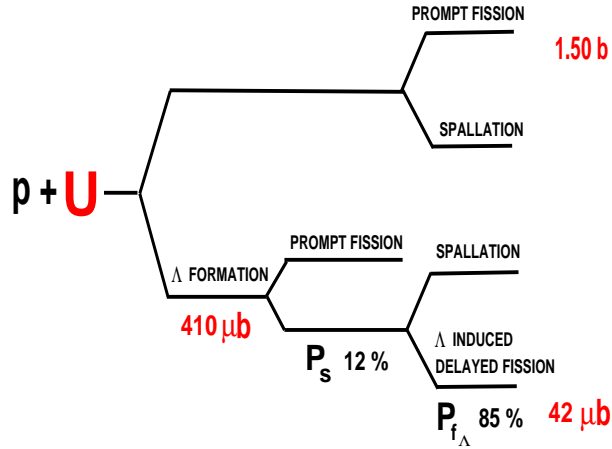,width=8cm,height=6cm,
        bbllx=18, bblly=387, bburx=605, bbury=716, clip=}
\caption{\small Schematic view of the relative population of the
delayed fission channel and competing reaction channels in p+U collisions.
Here P$_s$ denotes the average survival probability of the
hypernuclei against prompt fission while P$_{f_{\Lambda}}$ describes the probability of
fission induced by the delayed $\Lambda$N $\rightarrow$ NN decay.
}
\label{fig:fission}       
\end{figure}

The hypernuclei have been produced in interactions
of protons from the COSY - J\"ulich synchrotron at
an energy T$_{lab}$=1.9 GeV with a U target.  A very thin target has
been used such  that the  hypernuclei - produced with some momentum distribution -
can move out of the target area practically without distortions. Now the detection of delayed fission
fragments, {\it i.e. }
the position distribution of delayed fission events,
allows to extract information on the lifetime of the hyperons if the velocity of the hypernuclei
in the laboratory is known.
It should be emphasized that, at the energy of 1.9 GeV,
proton - nucleus collisions lead to
various processes other than  delayed fission, {\it e.g.} spallation,
fragmentation, or -- most importantly -- a prompt fission
of uranium nuclei.
The relative cross sections and survival probabilities are shown schematically in fig. \ref{fig:fission}.

%

The detection of hypernuclei and a measurement
of their lifetime via an observation of delayed fission events
in the presence of prompt fission events, that are more abundant by about five orders of
magnitude, is a quite difficult task.
This problem has been solved by an application of the recoil
distance method
as proposed by Metag et al. \cite{MET74} for
measurements of the lifetime of long living fission isomers in the
presence of a large background of prompt fission events.

\begin{figure}[h]
%
\centering
\epsfig{file=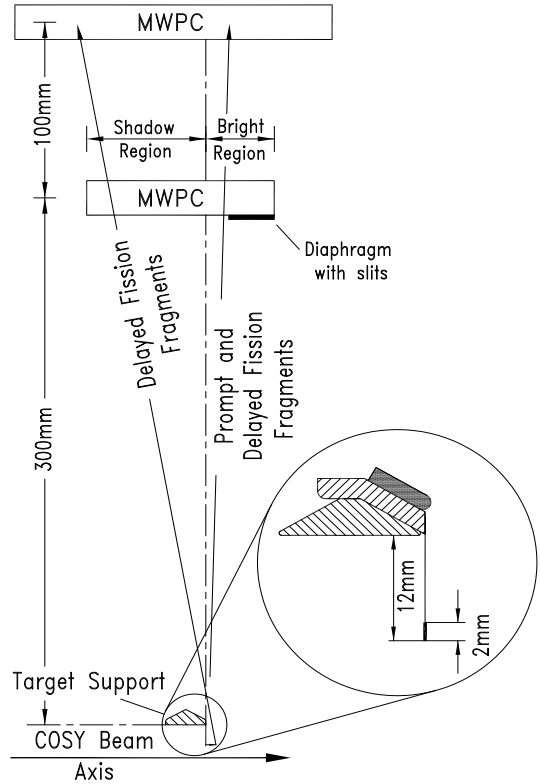,width=8cm,height=11cm,
        bbllx=0, bblly=0, bburx=555, bbury=801, clip=}
%
\caption{\small Schematic view of the experimental setup and illustration
of the recoil distance method (see text for a description). The target holder and the shadow edge (black)
are shown on a magnified scale.}
\label{fig:fig1m}       
\end{figure}

A schematic view of the experimental setup is
shown in Fig. \ref{fig:fig1m}.
Two position sensitive detectors -- operating in coincidence --
are placed above the target parallel to the beam.
Low pressure multiwire proportional chambers ( MWPCs ) have been
chosen since they are sensitive to heavy fragments only.
The MWPCs allowed, besides a measurement of the position of hits, also to
determine the time - of - flight (TOF) of the fragments
between the detectors as well as the energy loss of the fragments.
These  detectors were partly screened by the target holder (cf.
magnification in Fig. \ref{fig:fig1m}) from the prompt
fission fragments which directly emerge from the target.
Thus, as will be discussed below, the shadowed (left) region could be irradiated
only by delayed fission fragments or long living isomers.
Furthermore, to prevent an overloading of the bright (right) part of the detectors
by prompt fission fragments, a diaphragm was placed below the
bright region of the lower MWPC.  Two slits, parallel to the beam direction,
were cut in the diaphragm to allow for the registration of a reduced
yield from prompt fission fragments.

We now come back to the physical background processes in the p+U reaction at
1.9 GeV. When a proton impinges on an uranium nucleus,
the latter undergoes prompt fission with a cross section of $\sim$1.5 barn
\cite{HUD76,VAI81}.
The created fission fragments can
a) hit the bright part of a position sensitive detector (denoted by
"MWPC" in fig. \ref{fig:fig1m}) or
b) hit the shadow part of the target holder (cf. magnification in fig. \ref{fig:fig1m})
and be stopped;  some of the fragments, however,
may scatter on the edge of the target holder and move
into the shadowed part of the detectors. This process was found to give the dominant
 background in the measurements to be presented below.

On the other hand, also hypernuclei can fission promptly in the target
giving a contribution to  the bright part of the detectors, only. Such
events are of no interest for our lifetime measurements.
The "cold" hypernuclei (after deexcitation by nucleon and $\gamma$ emission), that have
survived the prompt fission stage,
remain stable up to  the time when the hyperon decays.
The excitation energy released in the $\Lambda$ decay can induce 
a delayed fission
of the residual nucleus.
Since the hypernuclei move approximately in beam direction -- due to the momentum transferred from the
impinging proton --
the latter fission occurs in some distance from the target. Thus
the delayed fission fragments can hit both,
the bright (right)  and shadowed (left) part
of the detectors (cf. fig. \ref{fig:fig1m}). 
The position distribution of hits on the surface of
the shadowed part of the detector, measured from the
shadow edge, then is just a magnified distribution of the distance
between the target and the 
points at which the delayed fission occurs.
%
%
%
%
%
Thus the shape of the distribution  in the
shadow region
contains the information on the lifetime of the
$\Lambda$ - hyperon folded with the velocity distribution of the
hypernuclei.

The distance between
the target and the target holder, which defines the shadow edge,
is about a factor 27 smaller than the
distance from the target holder to the detectors. This leads to a
significant magnification
 of the position scale on to the detector
and, therefore, allows for an accurate measurement
of the position distribution of  fissioning hypernuclei.

In the actual experiment thin (30 $\mu$g/cm$^2$) uranium targets -- in the
form of UO$_2$ backed on a carbon foil of 28 $\mu$g/cm$^2$
thickness --
were placed in the internal proton beam.  The details
of the target construction as well as other experimental techniques
are described in ref. \cite{PYS99}.

 In order to prove that the MWPCs
detect the fission fragments from delayed fission
of hypernuclei in the shadow region, but not lighter background
particles,
the following tests were performed:

\begin{itemize}

\item
The MWPCs were irradiated with minimum ionizing particles
($\beta^-$); it was found that the counting
efficiency for such particles is below 10$^{-11}$.

\item
A pure carbon foil was used as a target,  leaving all the other experimental
setups unchanged; the measured spectra in the shadowed part
of the detectors were found to be empty.

\item
A $^{252}$Cf source was placed in the target position and two--dimensional
energy loss ($\Delta E$) versus time - of - flight (TOF) spectra (between the two MWPCs)
were recorded; the measured $\Delta E - TOF$ distributions were found to be in full agreement
with Monte Carlo calculations taking into account the mass, charge and velocity of fragments
according to the Viola systematics \cite{VIO85} for the fission of californium.

\end{itemize}

Another type of background is expected from hypernuclei, which undergo
prompt fission in the target.  The hyperon, which is bound
in one of the fission fragments, decays in flight and can "kick" the
fragment towards the shadow region. However,
it has been shown by Monte Carlo simulations  that these hyperfragments
can hit the
shadowed region of the detectors only in a very narrow region
of 1--2 mm close to the edge
of the shadow region and, thus, do not contribute to the distribution
which was used for the extraction of the lifetime of hypernuclei.

In this experiment special precautions have  been taken to decrease
the background, which mainly stems from prompt fission fragments
that
scatter at the edge of the target holder or in the entrance foils
of the multiwire proportional counters:

The edge of the target holder, which served as a diaphragm -- defining
the shadowed region of the detector as shown in fig. \ref{fig:fig1m} 
(the dark dashed part of the holder) 
-- was shaped cylindrically
in the present experiment.
This geometry assured that the prompt fission
fragments, which enter into the material of the holder at larger angles
with respect to the direction perpendicular to the beam,
were absorbed in the holder and could not contribute
to scattering into the shadowed region.
This modified shape was found to be particularly important for measurements with an
uranium target since the prompt
fission cross section  reaches $\sim$ 1.5 barn,
which is $\sim 36000$ times larger than the delayed fission cross section (cf. fig. 1).

A new data acquisition system has been applied allowing for
the acceptance  of
a higher event rate than in the previous experiments. This enabled
us to  collect a higher number of delayed fission events before
the  internal structure of the thin target became distorted.
Furthermore, the
shape of the target was continuously recorded by a TV camera and photos
of the target were stored every several minutes.  This allowed for
corrections in the {\it off-line } analysis, which were necessary
because the uranium targets changed their shape during irradiation.

The background of the position distribution of events in the shadow
region of the detectors has been estimated performing
the measurements at T$_p$=1.0 GeV, which is much lower than the
threshold energy for the $\Lambda$-hyperon production in nucleon-nucleon
collisions.
Thus at this energy the cross section for hypernucleus formation is
negligible ($\sim$ 4 orders of magnitude smaller than at T$_p$=1.9 GeV)
whereas the prompt fission yield is about the same.
The COSY accelerator
was operated in the supercycle mode, {\it i.e.} there were three cycles
(each of $\sim$ 18 s duration) of  acceleration  and irradiation 
of the target;  two of them at
the higher energy of 1.9 GeV and one at 1.0 GeV.  This allowed to study
the effect and the background
almost concurrrently, \emph{i.e.} for the same shape and thickness of
the target.

The distribution of hit positions of the fission fragments in
the surface of the detector were projected on to the beam direction.
These experimental distributions were then compared with
simulated distributions, which have been evaluated from the velocity
distribution of the hypernuclei and the lifetime of the $\Lambda$ - hyperon
in the hypernuclei.  In these simulations
the lifetime was treated as a free parameter
whereas the velocity distribution was taken
from a theoretical analysis performed in the framework
of the coupled channel
Boltzmann-Uehling-Uhlenbeck model \cite{RUD95,RUD96,Cass99}
(for the first -- fast stage of the reaction)
and the Hauser-Feshbach model (for the second -- slow reaction stage).
It was not possible to determine experimentally
the velocity distribution of heavy hypernuclei because this would involve coincidence
measurements for the fragments.
Such measurements -- in the presence of prompt fission fragments being  more abundant by about five
orders of magnitude --
would be completely obscured by  random coincidences.

Since the number of events in the position distributions was not very large,
a Poisson (instead of Gaussian) probability distribution $p(n_i)$
has been used to simulate
the number of counts $n_i$ for each position bin:
\begin{eqnarray}
\label{eq:1}
p(n_i) & = & \frac{\lambda_i^{n_i} \cdot \exp(-\lambda_i)}{n_i!} \,\,, \\
\label{eq:2}
\lambda_i(\tau) & = & \alpha(\tau) \cdot l_i(\tau) + b_i \,\,,
\end{eqnarray}
where $\lambda_i(\tau)$ is the expected number of counts in the bin $"i"$
for the detection of delayed fission events --
depending on the lifetime $\tau$.
In (\ref{eq:2}) $l_i(\tau)$ is the number of events from the simulated
(non normalized)
distribution , $\alpha(\tau)$ is a normalization factor to be found
from the fit procedure, whereas $b_i$ is the  number of counts
of the background.

The distributions measured at T$_p$=1.0 GeV were employed for an estimate
of the background in the shadowed region of the detector
(after normalization
to the same number of events in the bright part of the detector as
for T$_p$=1.9 GeV ).  The search for the normalization factor $\alpha(\tau)$
has been performed by means of the "maximum likelihood method", which amounts to
find a
value of $\alpha(\tau)$ that maximizes the logarithmic
likelihood function $L(\alpha)$,
\begin{equation}
\label{eq:3}
L(\alpha) = \ln(\, \Pi_i \,\, p(n_i;\alpha(\tau)) \,) ,
\end{equation}
for a given value of the lifetime $\tau$.  Then the best lifetime $\tau$ was
extracted by the same -- maximum likelihood -- prescription, which
allows also for an estimate of the error for $\tau$ as the
difference between
the solution for $\tau$ and its value $\tau'$ corresponding to the
likelihood function
\begin{equation}
L(\tau') = L(\tau) - \frac{1}{2}
\end{equation}
(see {\it e.g.} ref. \cite{PAR00}).

\begin{figure}
\centerline{
\epsfig{file=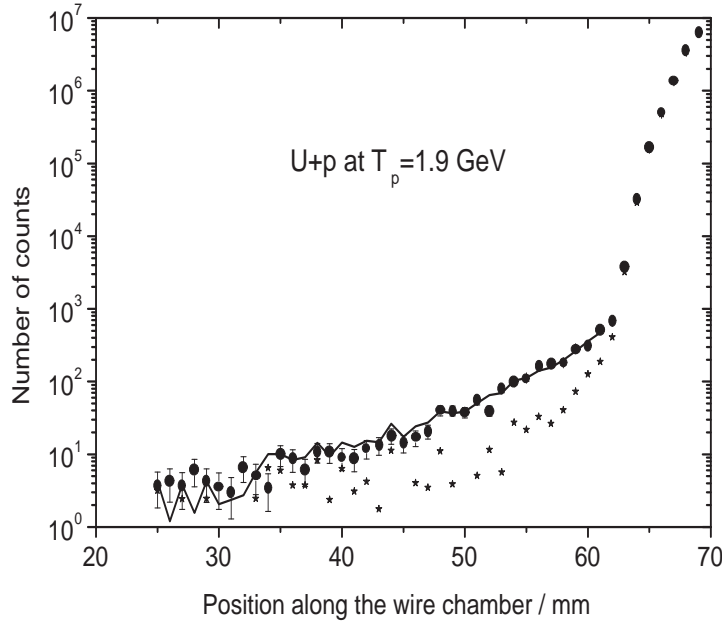,width=10cm,height=10cm,
        bbllx=0, bblly=0, bburx=300, bbury=250, clip=}}
\caption{ Experimental position distribution (full dots),
background (stars) and results of the
simulation (solid line). The fitted line is not smooth because
the background -- added to the theoretical distribution --
fluctuates due to small statistics.  The fit was performed
for the position range from 25 mm to 62 mm along the wire chamber.
For positions larger than 62 mm the experimental points are perfectly
reproduced by the background itself since the effect of delayed fission
is orders of magnitude smaller than that of the prompt fission.}
\label{fig:distr}
\end{figure}
The results of the fit are presented in Fig. \ref{fig:distr} by the solid line
in comparison to the data for T$_p$=1.9 GeV (full dots
with  error bars). The stars show the background events measured at
T$_p$=1.0 GeV and normalized to T$_p$=1.9 GeV data in the bright region
of the detector.

Within the analysis described above the following value for the
lifetime was found:
\begin{center}
$\tau_{\Lambda}$ = 138 $\pm$ 6 (stat.)   ps.
\end{center}

\section{Discussion of systematic errors}
\label{sec:discu}

The statistical error quoted in this result was determined
from the maximum likelihood method described above.  It only
depends on the statistics of the experimental position distributions.
The systematic errors in addition have various
origins; they emerge from

\begin{description}
\item[ 1.) ] uncertainties in the velocity distribution of hypernuclei,
\item[ 2.) ] an anisotropic emission of the fission fragments in their rest frame,
\item[ 3.) ] a not uniform irradiation of the target by the proton beam,
\item[ 4.) ] a modification of the position and shape of the targets
             during the measurements,
\item[ 5.) ] the background treatment,
\item[ 6.) ] the search procedure for the best lifetime.
\end{description}
These sources of errors will be separately discussed in more detail below.

\subsection{The velocity distribution of "cold" hypernuclei}
\label{subsec:velocity}

A velocity distribution in $v$ and a lifetime $\tau$
enter the simulations of the position
distributions via the product $v \cdot \tau$, \emph{i.e.}
the relative error
of the lifetime is approximately equal to the relative
error in the average velocity.
The velocity distributions in these investigations have been
taken from our theoretical calculations \cite{RUD98} that are based on the
transport code of Wolf et al. \cite{WOL90} and Maruyama et al. \cite{Tomo}.
Thus it is extremely important
to check whether these calculations are reliable with respect to the momentum
transfer of the proton to the residual nucleus.

\begin{figure}[h]
\centering
\resizebox{0.7\textwidth}{!}
{%
\epsfig{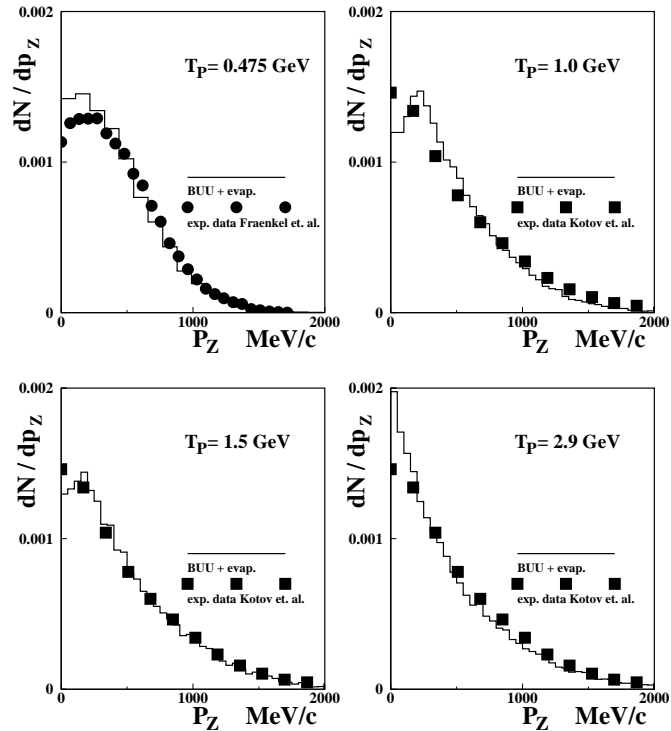}
}
\caption{ Comparison of the longitudinal momentum distribution for
p + $^{238}$U at $T_{lab}$ = 0.475 GeV, 1.0 GeV, 1.5 GeV, and 2.9 GeV
from the CBUU calculation (solid histograms) with the data
from Fraenkel et al. \cite{FRA90}(full dots) at 0.475 GeV and
Kotov et al. \cite{KOT74} at 1.0 GeV (full squares).  }
\label{fig:kotov}
\end{figure}

We show in fig. \ref{fig:kotov} the
longitudinal momentum distribution of the residual nuclei from our
coupled channel Boltzmann -- Uehling -- Uhlenbeck
(CBUU) + evaporation calculation (solid histogram)
for p + $^{238}$U at $T_{lab}$ = 475 MeV  \cite{RUD98} in
comparison to the data of Fraenkel et al. \cite{FRA90} (full
dots). Of course there is no hypernucleus formation at the
energy of 475 MeV.
However, when subtracting the kaon energy and
the nucleon-hyperon difference in mass for
the proton beam energy $T_{lab} \approx $ 1200
MeV, this should lead to similar kinematical conditions as for the
system considered.

In the energy range up to 3.0 GeV the  momentum transfer to the
residual nucleus in the reaction p + $^{238}$U was measured by
Kotov et al. \cite{KOT74}.
As pointed out by these authors the
average values for the momentum transfer show a very weak energy
dependence. Also in  fig. \ref{fig:kotov} the  CBUU + evaporation
calculations \cite{RUD98}  are presented for $T_{lab}$= 1.0, 1,5, 2.9 GeV and
compared to the experimental distributions (full squares)
measured by Kotov et
al.  \cite{KOT74} for $T_{lab}$ = 1.0 GeV.
It is evident that the calculations show a rather
weak energy dependence, too, and describe well the experimental
distributions. The good agreement with the data in this wide
kinematical regime demonstrates the accuracy of the
theoretical approach, which should be of the same quality when gating
on events with hypernucleus formation.

The reliability of our  CBUU  calculations
has been tested -- besides the analysis of prompt fission data -- also by
differential kaon spectra \cite{KUL00} since the kaons are emitted
in associated strangeness production reactions together with the hyperons.
Furthermore, by varying the hyperon-nucleon elastic cross section
\cite{RUD95,RUD96}
within a factor of two we verified that the impact on the
velocity distribution of hypernuclei is small, since the U target
leads to a large number of rescattering processes.
We found that the average velocity does
not change by  more than 3\% what leads to      $\sim$  2 ps for the standard
deviation of the lifetime.
%
%

%
\subsection{Anisotropic emission of the fission fragments}
\label{subsec:isotropic}

The angular distributions of fission products show a rather small
aniso\-tropy ($\approx$ 10 -- 20 \%)
even when gating on mass and charge  of fission
fragments as shown in refs. \cite{ZOL73,PAN81}.  A twofold
averaging is performed in our experiment, {\it i.e.} i) over
different fissioning hypernuclei and ii) over different fission
fragments.  Thus, it is expected that the assumption of isotropic
angular distributions is well justified.  The influence of possible
anisotropies  on the lifetime was estimated by
geometrical considerations to be approximately  2 ps.

\subsection{Non-uniform irradiation of the target}
\label{subsec:uniform}

The length of the active part of the target,
{\it i.e.} the UO$_2$ layer, was  about 2 mm, whereas the
full length of the target was 12 mm.
The assumption of a uniform irradiation of the active
part of the target is justified by the fact that the vertical
radius of the beam was about 3 mm. The
magnification of the position scale due to a different distance
from the active part of the target to the target holder and from
the target holder to the detector ($\approx$ 300 mm) thus was  on
average $\approx$ 27.
The assumption of a uniform irradiation of the active part of the
target is justified by the fact that the vertical radius of 
the beam was about 3 mm.
The difference between the average  magnification factor,
corresponding to a uniform
irradiation of the target,  and the smallest magnification, corresponding
to an irradiation of the lowest part of the target,
is $\approx$ 7\% which corresponds to a $\sim$10 ps deviation for the lifetime
obtained in the present experiment.
We note,  however, that the accelerated proton beam has
been moved upward on the target to achieve a
uniform counting-rate distribution over an irradiation cycle.
Taking this experimental boundary condition into account,
the error arising from a non-uniform irradiation of the target
in the present experiment is about 4 ps.

\subsection{Modifications of position and shape of the targets}
\label{subsec:shape}

The uranium targets in form of UF$_4$ (as
used  in the previous experiments) and UO$_2$ (in the present
work) were mechanically not very stable, i.e. they changed their
shape during irradiation.  Two high voltage electrodes were placed
closely to the target and  used to straighten it.
Due to the continuous control of the position and the shape of the
target it was possible to introduce corrections in the off-line
analysis, if the position or shape of the target  had been changed.
In the first case, {\it i.e.} a change of the target position, such
a correction was done in shifting the scale of the position
distribution.  This procedure involved a negligibly small error.
The following methods were used
to correct for the change of the target shape; i)
the simulated distributions were calculated for several observed
shapes of the target and the resulting theoretical distributions
were averaged over the shapes to simulate the experimental
distribution, or ii) the average shape of the target was taken for
the simulation.
Both methods of correction lead to almost the same values of the
lifetime and the inaccuracy involved in this procedure was
$\approx$ 4 ps.

\subsection{Background treatment}
\label{subsec:background}

  In the present work the data measured at T$_p$ = 1.0 GeV served --
after normalization of the counts to the 1.9 GeV data in the bright
part of the detectors -- to
estimate the background in the shadow region of the detectors.
It was found that the inaccuracy of background estimation, as
arising from the small statistics of the data at 1.0 GeV,
might introduce an error of about 3 ps for the lifetime.

\subsection{Search procedure for the lifetime}
\label{subsec:search}

  The fitting procedure of the simulated distributions
to the experimental distribution could be performed either with
the minimum $\chi^2$ - method (normally related to the assumption
of Gaussian statistics for the number of events in the individual
position bins) or with the maximum likelihood method (which is not
limited to Gaussian statistics and can be used also for a Poisson
distributions of events).  It was checked that both methods work
well in the case of high statistics data and lead to the same value for
the lifetime. However, for a small number of events the maximum
likelihood method with a Poisson distribution of events is
preferable \cite{ZYC99}. In this case the reanalysis of the low
statistics data \cite{OHM97} led to a  significant modification
(see Table \ref{tab:1}).

The other problem, which appears when fitting the simulated
distribution to the experimental one, is related to the position
range of the detector chosen for the fit procedure.  Neglecting
the position bins, that are placed closely to the shadow edge,
decreases significantly the statistics of events.  However, these
bins are  most strongly affected by the background originating
from  small angle scattering of prompt fission fragments on the
diaphragm as discussed above.  Thus, some compromise had to be reached, which was
most difficult in the case of uranium targets, that are plagued by the
largest background (as compared to Au and Bi targets).
With good statistics data  both  problems  introduce a rather
small error of 2 ps.

In summary, the systematic errors discussed above sum up
for the present experiment to  17 ps.

\section{Summary \& Conclusions}
\label{sec:summa}

The present work reports results on the lifetime measurement of
very heavy hypernuclei produced in p+U collisions.
%
%
%
\begin{table}[h]
\caption{ The lifetime of heavy hypernuclei measured at
COSY-J\"ulich with a proton beam and uranium targets. }
\label{tab:1}
\begin{tabular}{lll}
\hline\noalign{\smallskip}
 $\tau_{\Lambda}$/ps & Ref. & Comment   \\
\hline\noalign{\smallskip}
\hline\noalign{\smallskip}
240$\pm$60 ps & \cite{OHM97} & low statistics,\\
              &              & Gaussian distribution in the number of events, \\
              &              & $\chi^{2}$  fit \\
\hline\noalign{\smallskip}
194$\pm$55    & \cite{ZYC99} & reanalysis of the data from \cite{OHM97},\\
              &              & Poisson distribution in the number of events, \\
              &              & maximum likelihood method \\
\hline\noalign{\smallskip}
239$\pm$26 ps & \cite{PYS98} & moderate statistics, several targets,\\
              &              & Gaussian distribution in the number of events, \\
              &              & $\chi^{2}$  fit \\
\hline\noalign{\smallskip}
218$\pm$35 ps & \cite{ZYC99} & reanalysis of the data from \cite{PYS98}, \\
              &              & Poisson distribution in the number of events, \\
              &              & maximum likelihood method \\
\hline\noalign{\smallskip}
152$\pm$10(stat.)$\pm$25(syst.)& \cite{KUL00} & moderate statistics, \\
                               & & large background, \\
                               & & Poisson distribution in the number of events, \\
                               & & maximum likelihood method  \\
\hline\noalign{\smallskip}
138$\pm$6(stat.)$\pm$17(syst.) & present & good statistics,\\
                        &   work & Poisson distribution in the number of events, \\
                               & & maximum likelihood method \\
\hline\noalign{\smallskip}
\end{tabular}
\end{table}
%
%
This lifetime is determined almost completely by the
nonmesonic decay of the $\Lambda$-hyperon in the heavy
hypernucleus. The lifetime $\tau_{\Lambda}$ obtained in the present
experiment is compared in Table \ref{tab:1} with the results of
previous experiments performed by the COSY-13 collaboration  using
the internal proton beam  at COSY-J\"ulich and an uranium target.

The statistical error of the present data is significantly smaller
than those obtained in all previous experiments with uranium
targets because the statistics achieved in the present experiment is
the best (around 3000 events as compared with several hundreds
events in ref. \cite{PYS98,KUL00} and less than two hundred events in
ref. \cite{OHM97})

The errors given in ref. \cite{ZYC99} were evaluated by adding
squares of the statistical and systematic errors.  Using this
method of error presentation the result of Kulessa et al.
\cite{KUL00} has an error of 27 ps and the lifetime measured in
the present work has a total error of 18 ps.  Both  values are
significantly smaller than the errors of the previous measurements,
i.e. 55 ps and 35 ps, respectively.
Since the present data are obviously the most accurate,
we recommend for future analysis  to use as the lifetime of very
heavy hypernuclei, {\it i.e.} A $>$ 200, the value from our present work:
\begin{center}
$\tau_{\Lambda}$ = [138 $\pm$ 6(stat.) $\pm$ 17(syst.)] ps .
\end{center}

This result agrees very well with the lifetime found
from experiments using antiproton annihilation on  U targets
\cite{BOC86}, \cite{ARM93} where
 $\tau_{\Lambda}$ = 130 $\pm$ 30(stat.) $\pm$ 30(syst.)] ps
was quoted.
Since the energy involved in antiproton annihilation
is similar to the transfered energy in the present experiment, which studied
p+U collisions at T$_p$=1.9 GeV,
one can expect that roughly the same (A,Z) range of hypernuclei
was produced in both experiments.

We point out that
proton induced hypernucleus production has a couple of
advantages in comparison to antiproton annihilation: i) it was possible
to obtain a larger statistics of events, \emph{i.e.} a
smaller statistical error;
ii) due to a larger momentum transfer in proton induced collisions the hypernuclei
decay at a larger distance from the target than in the antiproton experiment, which
improves the spatial resolution;
iii) one can vary the kinetic energy of protons, \emph{i.e.} switching
on and off the production of hypernuclei, which is not possible
in antiproton induced strangeness production
since even for antiproton annihilation at rest  the center-of-mass energy is
above the strangeness production threshold.

The above points enabled us to more accurately observe the position distributions and
thus -- apart from a better statistics -- to reduce also the systematic error.

%
%

\end{document}